\documentclass[aps,prc,twocolumn,superscriptaddress,floatfix,reprint]{revtex4-1}
\usepackage{textcomp}
\usepackage{nicefrac}
\usepackage{longtable}
\usepackage{graphicx}
\usepackage{multirow}
\usepackage{amsmath}
\usepackage{amssymb}
\usepackage{makecell}
\usepackage{xcolor}

\begin{document}
 
 \title{Status of the $^{24}$Mg($\alpha,\gamma$)$^{28}$Si reaction rate at stellar temperatures}
 
 \author{P. Adsley}
 \email{philip.adsley@wits.ac.za}
 \affiliation{School of Physics, University of the Witwatersrand, Johannesburg 2050, South Africa}
 \affiliation{iThemba Laboratory for Accelerator Based Sciences, Somerset West 7129, South Africa}
 \affiliation{Institut de Physique Nucl\'{e}aire d'Orsay, UMR8608, IN2P3-CNRS, Universit\'{e} Paris Sud 11, 91406 Orsay, France}

\author{A. M. Laird}
\email{alison.laird@york.ac.uk}
\affiliation{Department of Physics, University of York, Heslington, York, YO10 5DD, United Kingdom}

\author{Z. Meisel}
\email{meisel@ohio.edu}
\altaffiliation{Affiliated with the Joint Institute 
for Nuclear Astrophysics Center for the Evolution of the Elements}
\affiliation{Institute of Nuclear and Particle Physics, Department of Physics \& Astronomy, Ohio University, Athens, Ohio 45701, USA}
 \date{\today}
 
\begin{abstract}

\begin{description}
\item[Background] The $^{24}$Mg($\alpha,\gamma$)$^{28}$Si reaction influences the production of magnesium and silicon isotopes during carbon burning and is one of eight reaction rates found to significantly impact the shape of calculated X-ray burst light curves. The reaction rate is based on measured resonance strengths and known properties of levels in $^{28}$Si.
\item[Purpose] It is necessary to update the astrophysical reaction rate for $^{24}$Mg($\alpha,\gamma$)$^{28}$Si incorporating recent modifications to the nuclear level data for $^{28}$Si, and to determine if any additional as-yet unobserved resonances could contribute to the $^{24}$Mg($\alpha,\gamma$)$^{28}$Si reaction rate. 
\item[Methods] The reaction rate has been recalculated incorporating updated level assignments from $^{28}$Si($\alpha,\alpha^\prime$)$^{28}$Si data using the {\tt RatesMC} Monte-Carlo code. Evidence from the $^{28}$Si($p,p^\prime$)$^{28}$Si reaction suggests that there are no further known resonances which could increase the reaction rate at astrophysically important temperatures, though some resonances do not yet have measured resonance strengths.
\item[Results] The reaction rate is substantially unchanged from previously calculated rates, especially at astrophysically important temperatures. However, the reaction rate is now constrained to better than 20\% across the astrophysically-relevant energy range, with 95\% confidence. Calculations of the X-ray burst lightcurve show no appreciable variations when varying the reaction rate within the uncertainty from the Monte-Carlo calculations.
\item[Conclusion] The $^{24}$Mg($\alpha,\gamma$)$^{28}$Si reaction rate, at temperatures relevant to carbon burning and Type I X-ray bursts, is well constrained by the available experimental data. This removes one reaction from the list of eight previously found to cause variations in X-ray burst light-curve calculations.
\end{description}
\end{abstract}

\maketitle

\section{Astrophysical Motivation}

The $^{24}$Mg($\alpha,\gamma$)$^{28}$Si reaction plays a role in stellar environments, namely in X-ray bursts, during carbon burning in massive stars and in neon burning, at temperatures from 0.5 to 2 GK. In the case of Type I X-ray bursts, recent studies by Cyburt {\it et al.} \cite{Cyburt_2016} (using the CF88 rate of Caughlan and Fowler \cite{CAUGHLAN1988283}) and  Meisel {\it et al.} \cite{Meisel_2019} (using the rate of Strandberg \cite{PhysRevC.77.055801}) have shown that the burst composition in the A=24, 28-30 region and the resulting light-curve are both sensitive to the $^{24}$Mg($\alpha,\gamma$)$^{28}$Si reaction rate. This reaction is influential at temperatures between about 0.5 GK (Gamow window: $E_r = 700$ to $1220$ keV \cite{PhysRevC.81.045807}) and 1.0 GK (Gamow window: $E_r = 1010$ to $1710$ keV \cite{PhysRevC.81.045807}). In particular, an increase in the $^{24}$Mg($\alpha,\gamma$)$^{28}$Si reaction rate by a factor of ten from the default rate modifies the light-curve convexity, a measure of the shape of the rise of the light-curve, by 25\% \cite{Meisel_2019}. Decreasing the $^{24}$Mg($\alpha,\gamma$)$^{28}$Si reaction rate, however, had a much smaller impact \cite{Cyburt_2016}.  

The light-curve from X-ray bursts may, by comparison to models, be used to extract neutron-star properties such as mass and radius, as well as the accretion rate. However, the models and thus the neutron-star data extracted are sensitive to the thermonuclear reaction rates used. Once reaction rates are well constrained then their potential influence on the light-curve may be minimised, reducing the uncertainty in the extraction of the neutron-star properties. Progress in constraining neutron star properties is particularly timely given the recent observation of a neutron star - neutron star merger (GW170817) and the detection of strontium in the resulting kilonova light curve \cite{Wat19}.

In massive stars, $^{24}$Mg is produced during carbon burning via the $^{20}$Ne($\alpha,\gamma$)$^{24}$Mg reaction following the conversion of two $^{12}$C nuclei into $^{20}$Ne by the reaction chains $^{12}$C($^{12}$C,$\alpha$)$^{20}$Ne and $^{12}$C($^{12}$C,$p$)$^{23}$Na($p,\alpha$)$^{20}$Ne. $^{24}$Mg is subsequently destroyed by neutron- or $\alpha$-particle capture making $^{25}$Mg and $^{28}$Si respectively. The abundances of magnesium and silicon isotopes depend on, amongst other factors, the relative strengths of the capture reactions onto $^{24}$Mg within the relevant temperature range of 1 to 1.4 GK associated with carbon-shell burning \cite{PhysRevC.77.055801}.

The factor of ten increase in the reaction rate in Refs. \cite{Cyburt_2016,Meisel_2019} was chosen as a plausible uncertainty for a nuclear reaction rate involving a relatively high level-density compound nucleus. However, there are existing evaluations of the $^{24}$Mg($\alpha,\gamma$)$^{28}$Si reaction rate from the STARLIB collaboration \cite{0067-0049-207-1-18} and from experimental studies \cite{PhysRevC.77.055801} which have much smaller uncertainties. Some of the potential causes of an increase in the reaction rate include the presence of additional unobserved resonances, changes in level assignments for resonances, systematic biases in measurements as well as systematic biases in the evaluation of the reaction rates resulting from e.g. mass evaluations. The purpose of this paper is to consider and quantify these potential sources of systematic or unaccounted uncertainties in the reaction rate. In order to do this, we briefly discuss the available nuclear data on $^{28}$Si and show that there are unlikely to be hidden systematic uncertainties in the reaction rate; demonstrate, using data from the $^{28}$Si($p,p^\prime$)$^{28}$Si reaction, that there are no unobserved additional resonances which could modify the $^{24}$Mg($\alpha,\gamma$)$^{28}$Si reaction rate at astrophysically relevant temperatures; and show that the mistaken level assignments of $^{28}$Si cannot result in significant changes to the reaction rate, leading to the conclusion that the uncertainties in the reaction rate result in no observable variation in the lightcurve of X-ray bursts.

\section{Nuclear Data}

The available nuclear data above the $\alpha$-particle threshold at 9.984 MeV up to $E_x = 12$ MeV ($E_r = 2000$ keV) are summarised in Table \ref{tab:Si28_nuclear_data}. The sources of resonance strengths are the direct measurements performed by Smulders and Endt \cite{SMULDERS19621093}, Lyons \cite{LYONS196925} and Strandberg {\it et al.} \cite{PhysRevC.77.055801}. Spectroscopic information is available from the $\gamma$-ray spectroscopy data obtained in $^{27}$Al($p,\gamma$)$^{28}$Si and $^{24}$Mg($\alpha,\gamma$)$^{28}$Si reactions of Brennesien {\it et al.} \cite{BrenneisenI,BrenneisenII,BrenneisenIII}, the $^{28}$Si($\alpha,\alpha^\prime$)$^{28}$Si data of Adsley {\it et al.} \cite{PhysRevC.95.024319}, the $^{28}$Si($p,p^\prime$)$^{28}$Si data of Adsley {\it et al.} \cite{PhysRevC.97.045807}, and the $^{28}$Si($e,e^\prime$)$^{28}$Si data of Schneider {\it et al.}. We briefly summarise these experimental data and the information obtained from them below.

\subsection{Measurements of $^{24}$Mg($\alpha,\gamma$)$^{28}$Si resonance strengths}

Maas {\it et al.} \cite{MAAS1978213}, and Smulders and Endt \cite{SMULDERS19621093} studied the $^{24}$Mg($\alpha,\gamma$)$^{28}$Si reaction using a 10 cm x 10 cm NaI crystal, measuring strengths for resonances from $E_r = 3246$ keV ($E_\alpha = 3787$ keV) down to $E_r = 1311$ keV ($E_\alpha = 1530$ keV). Smulders and Endt additionally performed angular-correlation analyses on resonances observed in  $^{24}$Mg($\alpha,\gamma$)$^{28}$Si and $^{27}$Al($p,\gamma$)$^{28}$Si reactions in order to assign spins and parities. Both of these experimental studies measured yield curves with energy scans rather than only performing on-resonance measurements.

Lyons \cite{LYONS196925} measured yield curves from a maximum resonance energy of $E_r = 2317$ keV ($E_\alpha = 2703$ keV) down to the resonance at $E_r = 1164$ keV ($E_\alpha = 1358$ keV). Two NaI crystals in close geometry, functioning as a total-absorption spectrometer were used to measure the yields from the $^{24}$Mg($\alpha,\gamma$)$^{28}$Si reaction. The same experimental equipment had already been used to measure the $^{27}$Al($p,\gamma$)$^{28}$Si reaction \cite{LYONS19691}. In this experiment, resonances were observed down to $E_r = 1158$ keV. Lyons also performed yield curves by scanning the energy of the incoming beam.

he experimental data of Strandberg {\it et al.} \cite{PhysRevC.77.055801} scanned energies between $E_r = 1337$ keV ($E_\alpha = 1560$ keV) and $E_r = 909$ keV ($E_\alpha = 1060$ keV). More detailed data were taken at specific resonance energies focusing on known or potential natural parity states, chosen based on the existing data on excited states in $^{28}$Si. Data were taken in smaller energy steps to scan over the resonances in question. The $E_r = 1311$-keV resonance which had been observed by Smulders and Endt, Maas, and Lyons was remeasured, allowing potential systematic deviations between data to be identified. Resonance strengths were measured for resonances down to $E_r = 1010$ keV. Below this energy (corresponding to $E_\alpha = 1178$ keV) no resonances were observed but upper limits on the resonance strengths for all lower lying resonances were determined.

Four NaI(Tl) and one HPGe clover were used to detect $\gamma$ rays resulting from $^{24}$Mg($\alpha,\gamma$)$^{28}$Si reactions. For the weaker resonances the detectors were operated in coincidence mode with the NaI(Tl) crystals detecting the high-energy primary $\gamma$ ray, with the clover detecting the $E_\gamma = 1779$-keV transition from the first-excited state to the ground state. This allowed for an improved signal-to-background ratio for the weaker resonances and a more robust extraction of the resonance strengths. For the stronger resonances, the singles data taken with the clover detector could be used to assign branching ratios from the primary $\gamma$ rays. These stronger resonances were those at $E_r = 1010$, $1158$ and $1311$ keV.

Brenneisen {\it et al.} \cite{BrenneisenI,BrenneisenII,BrenneisenIII} measured the $^{24}$Mg($\alpha,\gamma$)$^{28}$Si ($E_\alpha = 1500-4000$ keV) and $^{27}$Al($p,\gamma$)$^{28}$Si ($E_p = 630-4850$ keV) reactions to assign spins and parities of levels in $^{28}$Si. Information on resonance strengths in the astrophysically important region were not given.

\subsubsection{Consistency of resonances strengths}
\label{sec:ConsistencyOfResonanceStrengths}

The weighted means of each of the resonance strengths are computed from the available nuclear data using:

\begin{equation}
 \bar{x} = \sum^N_{i=0} x_i/\sigma_i^2 / \sum^N_{i=0} 1/\sigma_i^2,
\end{equation}
where the $x_i$ are the data and the $\sigma_i$ are the corresponding uncertainties.

The percentage relative difference of the resonance strengths given by:
\begin{equation}
 R = 100\frac{(x_i - \bar{x})}{\bar{x}}.
\end{equation}

\begin{figure}
\includegraphics[width=\columnwidth]{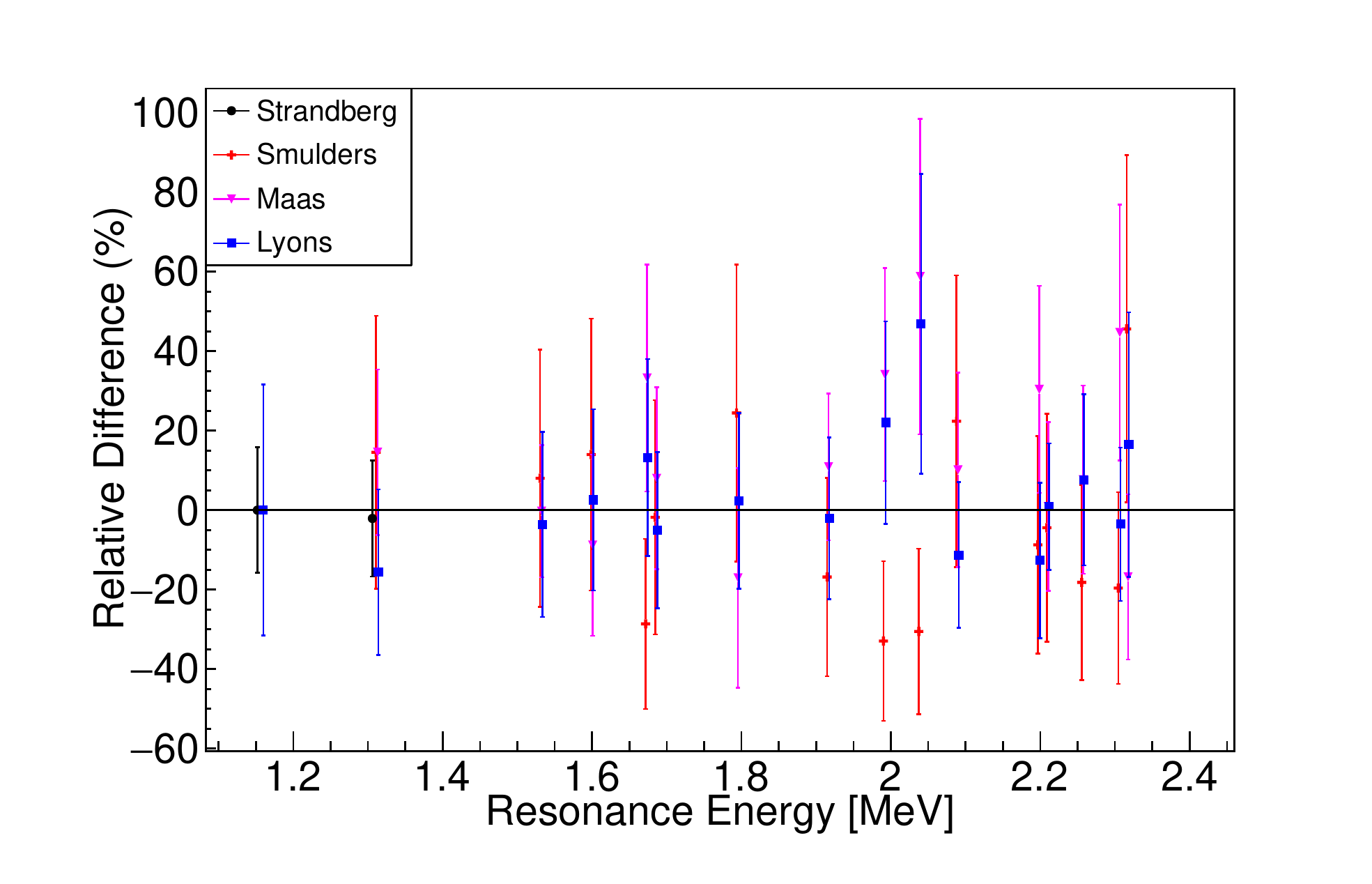}
 \caption{Relative difference (see the text for the definition) of resonance strengths from the data of Maas \cite{MAAS1978213}, Smulders and Endt \cite{SMULDERS19621093}, Lyons \cite{LYONS196925}, and Strandberg \cite{PhysRevC.77.055801}. }
 \label{fig:RelativeResonanceStrengths}
\end{figure}

The relative differences are shown in Figure \ref{fig:RelativeResonanceStrengths}. Note that only two points from the data of Strandberg {\it et al.} appear as only two resonances ($E_r = 1158$ keV and $1311$ keV) were measured in more than one experiment. In preparing this analysis, the uncertainties on all of the resonance strengths of Smulders and Endt \cite{SMULDERS19621093} were assumed to be 30\% of the reported value. Smulders and Endt report that the uncertainties in the absolute resonance strengths from their measurement range from 30\% for the strongest resonances to a factor of two for the weakest resonances, but it is unclear how these terms are defined or how the uncertainties vary between these two limits. We assume a 30\% uncertainty as this ensures that the errors are never over-estimated. 

The data of Strandberg {\it et al.} include both the resonance strengths determined in that experiment and adopted values which were obtained by scaling to the $E_r = 1311$-keV resonance strength from the Smulders and Endt measurement, though as these resonance strengths are consistent to within $1\sigma$ it is not clear that this procedure is necessary. The data plotted in Figure \ref{fig:RelativeResonanceStrengths} use the unscaled results of Strandberg {\it et al.} 

It is clear from Figure \ref{fig:RelativeResonanceStrengths} that the existing resonance strengths are entirely consistent with each other. In fact, given the small reduced $\chi^2$ values computed for the various resonances (for example. $\chi^2/N = 0.417629$ for the $E_r = 1157$-keV resonance) it appears that the systematic errors in these resonance strength measurements may be overestimated.

\subsection{The $^{28}$Si($\alpha,\alpha^\prime$)$^{28}$Si reaction}

The $^{28}$Si($\alpha,\alpha^\prime$)$^{28}$Si reaction has been measured at iThemba LABS in South Africa using the K600 at very forward angles including 0 degrees \cite{PhysRevC.95.024319} and at RCNP Osaka using the Grand Raiden magnetic spectrometer \cite{PhysRevC.93.064325}. In Ref. \cite{PhysRevC.95.024319}, the $\ell$-value, and therefore spin and parity, for the state were derived from the differential cross section. Following this experiment two changes to the spin and parity assignments of $^{28}$Si levels were made. First, the $E_x = 10.806$-MeV level assignment was changed  from $J^\pi = 2^+$ to $J^\pi = 0^+$ and, second, a $J^\pi = 0^+$ level at $E_x = 11.142$ MeV was observed in addition to a known $J^\pi = 2^+$ level at $E_x = 11.148$ MeV.

The focus of the Osaka experiment of Peach {\it et al.} \cite{PhysRevC.93.064325} was on the isoscalar giant resonances which lie at higher excitation energies than those relevant for the $^{24}$Mg($\alpha,\gamma$)$^{28}$Si reaction. While some low-lying discrete $\ell = 0$, $\ell = 1$ and $\ell = 2$ strength was observed in this experiment, no analysis of these states was performed and the energies of these narrow states are not given.

\subsection{The $^{28}$Si($p,p^\prime$)$^{28}$Si reaction at low energies}

Data on the $^{28}$Si($p,p^\prime$)$^{28}$Si reaction with the Q3D spectrometer at the Maier-Leibnitz-Laboratorium (MLL), Garching, Germany, were taken using a 18-MeV proton beam on a 40-$\mu$g/cm$^2$-thick $^{28}$SiO$_2$ target on a carbon backing. These data were published as a calibration spectrum and as evidence of the selectivity of the $^{28}$Si($d,d^\prime$)$^{28}$Si reaction to $\Delta T=0$ transitions as part of an experiment investigating levels in $^{26}$Mg \cite{PhysRevC.97.045807}.

\begin{figure}
 \includegraphics[width=\columnwidth]{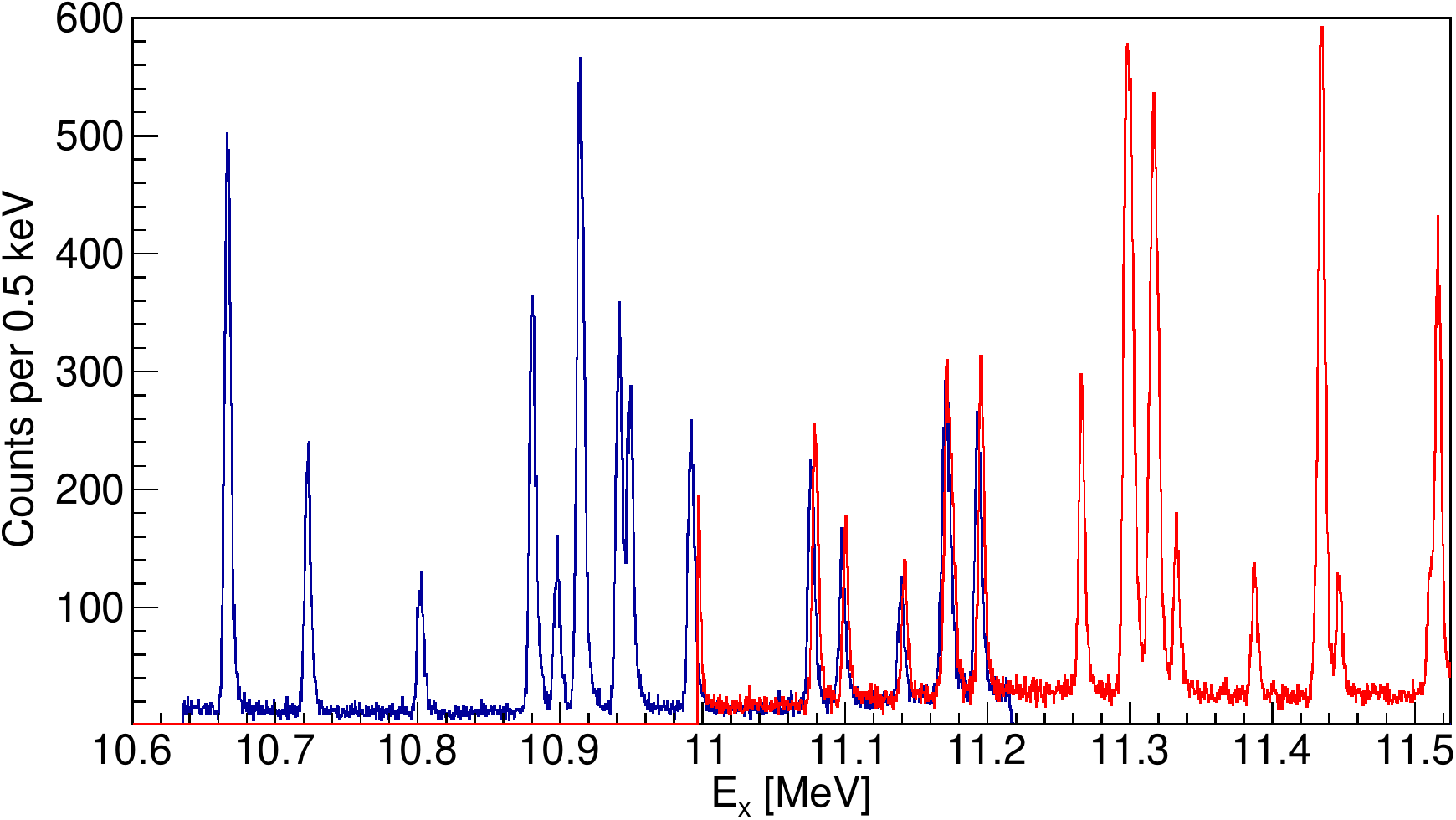}
 \caption{Excitation-energy spectrum from the $^{28}$Si($p,p^\prime$)$^{28}$Si reaction at field settings 1 (blue) and 2 (red).}
 \label{fig:Si28ppSpectrum}
\end{figure}

An excitation-energy spectrum from the $^{28}$Si($p,p^\prime$)$^{28}$Si reaction is shown in Fig. \ref{fig:Si28ppSpectrum}. Data were taken in two different exposures at different field settings with an overlapping region in the centre. Almost all peaks correspond to states in $^{28}$Si: the two levels at approximately $E_x = 11.3$ MeV are the $E_x = 11.080$- and $E_x = 11.097$-MeV states in $^{16}$O. The $E_x = 10.957$-MeV state in $^{16}$O is observed at around $E_x = 11.17$ MeV.

Importantly, proton inelastic scattering at these energies is non-selective \cite{doi:10.1139/p71-062,MOSS1976413,PhysRevC.89.065805} meaning that comparison with all known states from more selective reactions such as $^{24}$Mg($\alpha,\gamma$)$^{28}$Si and $^{27}$Al($p,\gamma$)$^{28}$Si may be made. The ($p,p^\prime$) reaction at these energies is not a resonant-scattering reaction and so the selectivity of the proton coupling of the states is limited.

All of the levels observed in the $^{28}$Si($p,p^\prime$)$^{28}$Si reaction of Ref. \cite{PhysRevC.97.045807} correspond to known levels in $^{28}$Si. The only levels which are not observed are those at $E_x = 11.148$ MeV and $E_x = 11.242$ MeV. However, as discussed in Section \ref{sec:NucDat}, it is likely that neither of these levels exists.

\subsection{The $^{28}$Si$(p,p^\prime$)$^{28}$Si reaction at high energies}

The $^{28}$Si($p,p^\prime$)$^{28}$Si reaction has also been performed at high energy with $E_p = 295$ MeV \cite{PhysRevLett.115.102501}. This is a very different reaction to the low-energy $^{28}$Si($p,p^\prime$)$^{28}$Si reaction in that it is selective to certain kinds of states and the resulting differential cross sections are indicative of the spin and parity of the populated state. While the published article \cite{PhysRevLett.115.102501} discussing these levels focusses on the quenching (or lack thereof) of the M1 transitions to $J^\pi = 1^+$ states in $N=Z$ $sd$-shell nuclei, the associated PhD thesis \cite{MatsubaraPhDThesis} reports differential cross sections and spin-parity assignments.

\subsection{The $^{28}$Si($e,e^\prime$)$^{28}$Si reaction}

Schneider {\it et al.} \cite{SCHNEIDER197913} have reported electron-scattering data from $^{28}$Si. These data focus on the population of $J^\pi = 1^+$ states in $^{28}$Si. In passing, they report $J^\pi = 2^+$ levels at $E_x = 10.515$, $10.807$, $11.148$, $12.072$, $12.439$ and $12.726$ MeV. Ref. \cite{SCHNEIDER197913} refers to another publication which will discuss these levels but this does not appear in the literature. However, at least two of these assignments are in conflict with spin-parity assignments made using other probes, as discussed below. Distinguishing $J^\pi = 0^+$ and $J^\pi = 2^+$ states with electron scattering is not possible using the longitudinal momentum distribution but only with the transverse momentum distribution \cite{PrivateCommunicationPvNC}. For this reason, confusing $J^\pi = 0^+$ and $J^\pi = 2^+$ assignments is a very present risk in the interpretation of inelastic electron scattering.

\section{Combined Nuclear Data}
\label{sec:NucDat}

Having discussed the various different sources of nuclear data for this reaction, we present the known information on states in $^{28}$Si and corresponding information on the resonance strengths of the $^{24}$Mg($\alpha,\gamma$)$^{28}$Si reaction in Table \ref{tab:Si28_nuclear_data}. Resonances for which level assignments have been changed since the results of Strandberg {\it et al.} \cite{PhysRevC.77.055801} or which differ from those listed in the ENSDF \cite{ENSDF} are discussed in more detail below.

\begin{table*}
\caption{Available nuclear data for states in $^{28}$Si between the $\alpha$-particle threshold and $E_r = 2000$ keV. For states above $E_r = 2000$ keV, refer to the STARLIB compilation of resonance strengths \cite{0067-0049-207-1-18}. This data table is based on Ref. \cite{BrenneisenIII} with additional level assignments from Ref. \cite{PhysRevC.95.024319} and ENSDF \cite{ENSDF}. For states without measured resonance strengths or experimentally determined upper limits, the resonance strength corresponding to the Wigner limit for the resonance is given in square brackets. The penultimate column indicates whether the level is included in the STARLIB default input file. The final column lists the source of the resonance strength; WL means that the maximum resonance strength is given by the Wigner Limit. All states without an excitation-energy uncertainty have uncertainties below 1 keV as per Ref. \cite{BrenneisenIII}.}
\label{tab:Si28_nuclear_data}
\begin{ruledtabular}
 \begin{tabular}{ c c c c c c c}
  $E_x$ [MeV] & $E_r$ [keV] & $J^\pi$ & $\omega\gamma$ [eV] & Comments & In STARLIB & Source of $\omega\gamma$ \\ \\
  \hline  \\
  $10.182$	&	$198$	&	$3^-$		&			[$3.8\times10^{-28}$]	&	& \checkmark & WL	\\
  $10.190$	&	$206$	&	$5^-$		&			[$1.9\times10^{-29}$]	&	&	    & WL\\
  $10.210$	&	$226$	&	$3^+$		&				& Unnatural parity 	&	\\
  $10.272$	&	$288$	&	$0^+$		&			& Isovector		&	\\
  $10.311$	&	$327$	&	$4^+$		&			[$1.4\times10^{-19}$]	&	& & WL	\\
  $10.376$	&	$392$	&	$3^+$		&			& Unnatural parity	&	\\
  $10.418$	&	$434$	&	$5^+$		&		 	& Unnatural parity	&	\\
  $10.515$	&	$531$	&	$2^+$		&			[$8.0\times10^{-11}$]	&	& \checkmark	\\
  $10.541$	&	$557$	&	$3^-$		&			[$7.4\times10^{-11}$]	&	& & WL	\\
  $10.596$	&	$612$	&	$1^+$		&			& Unnatural parity	&	\\
  $10.668$	&	$684$	&	$3^+(2^+)$	&			[$9.7\times10^{-8}$]	& \makecell{Resonance strength limit for a \\$J^\pi = 2^+$ assignment.}	& & WL	\\
  $10.669$	&	$685$	&	$4^+$		&			[$3.0\times10^{-9}$]	&	& & WL	\\
  $10.725$	&	$741$	&	$1^+$		&		& Unnatural parity		&	\\
  $10.778$	&	$794$	&	$1^+-5^+$	&			[$4.2\times10^{-6}$]	&	\makecell{For a $J^\pi = 2^+$ assignment.\\Upper limit for resonance strength:\\ $\omega\gamma < 2\times 10^{-6}$.} 	& & \cite{PhysRevC.77.055801}	\\
  $10.806$	&	$822$	&	$0^+$		&			[$1.2\times10^{-5}$]	& \makecell{State reassigned as $J^\pi = 0^+$ in Ref. \cite{PhysRevC.95.024319}.\\Upper limit for resonance strength:\\ $\omega\gamma < 2\times 10^{-6}$.}	& \checkmark & \cite{PhysRevC.77.055801}	\\
  $10.884$	&	$900$	&	$(2,3)$		&		$<2 \times 10^{-6}$ & \makecell{Limit on combined strength of all resonances\\ at and below the $E_r = 900$-keV resonance.}		& & \cite{PhysRevC.77.055801}	\\
  $10.900$	&	$916$	&	$1^+$		&			& Unnatural parity		&	\\
  $10.916$	&	$932$	&	$3^-$		&			$<1.4 \times 10^{-5}$ &		& \checkmark & \cite{PhysRevC.77.055801}	\\
  $10.945$	&	$961$	&	$4^+$		&			$<1.6 \times 10^{-5}$ & \makecell{Limit on combined $E_r = 932/969$-keV\\resonance strength.}	& & \cite{PhysRevC.77.055801}	\\
  $10.953$	&	$969$	&	$2^+$		&			$<1.6 \times 10^{-5}$ & \makecell{Limit on combined $E_r = 932/969$-keV\\ resonance strength.}	& \checkmark & \cite{PhysRevC.77.055801}	\\
  $10.994(3)$	&	$1010$	&	$1^-$		&		$2.3(6)\times 10^{-4}$ & $J^\pi$ assignment from $^{28}$Si($\alpha,\alpha^\prime$)$^{28}$Si \cite{PhysRevC.95.024319}	&	\checkmark & \cite{PhysRevC.77.055801} \\
  $11.078$	&	$1094$	&	$3^-$		&			$62(11)\times 10^{-5}$ &	& \checkmark & \cite{PhysRevC.77.055801} \\
  $11.101$	&	$1117$	&	$6^+$		&			[$1.7\times10^{-6}$]		& \makecell{Upper limit is Wigner limit,\\ experimental limit could be smaller.}	& & \cite{PhysRevC.77.055801}	\\
  $11.142$	&	$1158$	&	$0^+$		&			$0.0020(3)$ 	&	& \checkmark & \cite{PhysRevC.77.055801}	\\
  $11.196$	&	$1212$	&	$4^+$		&			$22(4)\times 10^{-5}$ &		& \checkmark \\
  $11.242(6)$     &       $1258(6)$ &                       &                                      & \makecell{From $^{27}$Al($d,n$)$^{28}$Si \cite{PhysRevC.32.394}. See note in text\\ about probable non-existence of this state.} & & \\
  $11.266$	&	$1282$	&	$3^-$		&					& Omitted but $\omega\gamma$ likely extremely small. &   & \cite{PhysRevC.77.055801}\\
  $11.296$	&	$1311$	&	$1^-$		&			$0.099(11)$ 	&		& \checkmark & \cite{PhysRevC.77.055801} \\
  $11.333$	&	$1349$	&	$6^+$		&					&		& \\
  $11.388$	&	$1404$	&			&					&	\makecell{Not observed in $^{24}$Mg($\alpha,\gamma$)$^{28}$Si,\\seen in $^{28}$Si($p,p^\prime$)$^{28}$Si data of Ref. \cite{MOSS1976413}.} 	& \\
$11.515$    & $1531$ & $2^+$ &  $0.058(14)$ & &  \checkmark & \cite{Sallaska_2013}\\
$11.584$        & $1600$ & $3^-$  & $0.045(10)$  & & \checkmark & \cite{Sallaska_2013} \\
$11.657$        & $1673$ & $2^+$  & $0.119(26)$ & & \checkmark & \cite{Sallaska_2013} \\
$11.669$        & $1685$ & $1^-$  & $0.29(6)$ & & \checkmark & \cite{Sallaska_2013} \\
$11.778$        & $1794$ & $2^+$  & $0.037(8)$ & & \checkmark & \cite{Sallaska_2013} \\
$11.899$        & $1915$ & $(2^+,3^-,4^+)$  & $0.053(11)$ & & \checkmark & \cite{Sallaska_2013} \\
$11.981$        & $1997$ & $(2^+,3^-,4^+)$  & $0.091(19)$ & & \checkmark & \cite{Sallaska_2013} \\
 \end{tabular}
 \end{ruledtabular}
\end{table*}

\subsection{The $E_x = 10.806$-MeV state}

This resonance was reassigned from $J^\pi = 2^+$ to $J^\pi = 0^+$ in the $^{28}$Si($\alpha,\alpha^\prime$)$^{28}$Si measurement of Adsley {\it et al.} \cite{PhysRevC.95.024319}. This reassignment has limited impact on the reaction rate for two reasons: first, the single-particle limit for the resonance strength in these two cases is nearly identical and second, a more stringent limit on the reaction rate from the direct measurement of Strandberg {\it et al.} \cite{PhysRevC.77.055801} exists. This reassignment therefore has no impact on the rate.

A state at $E_x = 10.807$ MeV was observed in the $^{28}$Si($e,e^\prime$)$^{28}$Si data of Schneider {\it et al.} \cite{SCHNEIDER197913} and given a $J^\pi = 2^+$ assignment. However, it appears that this assignment is incorrect given the observation of the $J^\pi = 0^+$ state in the $^{28}$Si($\alpha,\alpha^\prime$)$^{28}$Si data of Adsley {\it et al.} \cite{PhysRevC.95.024319} and of a $J^\pi = 0^+$ state in the $^{28}$Si($p,p^\prime$)$^{28}$Si data of Matsubara {\it et al.} at $E_p = 295$ MeV \cite{PhysRevLett.115.102501,MatsubaraPhDThesis}. This mistaken assignment is probably due to the difficult in distinguishing $J^\pi = 0^+$ and $J^\pi = 2^+$ assignments with electron scattering \cite{PrivateCommunicationPvNC}.

\subsection{The $E_x = 10.994$-MeV state}

This level was listed as $J^\pi = (1,2^+)$ prior to the $^{24}$Mg($\alpha,\gamma$)$^{28}$Si experiment of Strandberg {\it et al.} \cite{PhysRevC.77.055801}. Observation of the resonance in the $^{24}$Mg($\alpha,\gamma$)$^{28}$Si reaction rules out a $J^\pi = 1^+$ assignment. The $^{28}$Si($\alpha,\alpha^\prime$)$^{28}$Si data of Adsley {\it et al.} \cite{PhysRevC.95.024319} gives a clear $J^\pi = 1^-$ assignment for this state. 

\subsection{The $E_x = 11.142$-MeV state}

The $E_x = 11.142$-MeV state was observed in $\gamma$-ray decays following $^{24}$Mg($\alpha,\gamma$)$^{28}$Si and $^{27}$Al($p,\gamma$)$^{28}$Si reactions \cite{BrenneisenI,BrenneisenII,BrenneisenIII}. As assignment of $J = 0-2$ was made on the basis of the feeding from a $J^\pi= 1^+$ excited state. This level has a clear $J^\pi = 0^+$ assignment in the $^{28}$Si($\alpha,\alpha^\prime$)$^{28}$Si measurement of Adsley {\it et al.} \cite{PhysRevC.95.024319}. This $J^\pi = 0^+$ assignment is supported by high-energy $^{28}$Si($p,p^\prime$)$^{28}$Si data of Matsubara {\it et al.} \cite{PhysRevLett.115.102501,MatsubaraPhDThesis} at $E_p = 295$ MeV who, additionally, do not report a $E_x = 11.148$-MeV state. 

A $E_x = 11.148$-MeV, $J^\pi = 2^+$ state was assigned in $^{28}$Si($e,e^\prime$)$^{28}$Si reactions at the S-DALINAC at Darmstadt \cite{SCHNEIDER197913}. Details as to the assignment of the spins and parities of the levels in this experiment are not given as the focus of the paper is on the $J^\pi = 1^+$ levels. Ref. \cite{SCHNEIDER197913} states that there will be a following paper which discusses these level assignments but this does not appear to be available. 

In the low-energy $^{28}$Si($p,p^\prime$)$^{28}$Si data presented in the present paper, only one level between $E_x = 11.14$ and $11.15$ MeV is observed (see Figure \ref{fig:Si28ppSpectrum}) and we assume that this is the $J^\pi = 0^+$ state seen in intermediate-energy scattering measurements. The $J^\pi = 2^+$ assignment based on the $^{28}$Si($e,e^\prime$)$^{28}$Si data is potentially problematic as has already been seen in the case of the $E_x = 10.806$-MeV level. We suggest that the assignment of a $J^\pi = 2^+$ level based on the $^{28}$Si($e,e^\prime$)$^{28}$Si reaction should be treated as tentative until confirmed through another source.

A $J^{\pi}=0^{+}$ assignment is in tension with the $^{24}{\rm Mg}(\alpha,\gamma)^{28}{\rm Si}$ results of Strandberg et al., who observe a 10\% branch from this resonance to the $J^\pi = 4^+$ level at $E_x = 4.62$ MeV \cite{PhysRevC.77.055801}. However, the state at this energy in the $^{28}$Si($e,e^\prime$)$^{28}$Si data of Schneider {\it et al.} \cite{SCHNEIDER197913} is strongly populated, implying a large transition strength linking this state and the ground state. For a $J^\pi = 2^+$ assignment, this disagrees with the branching ratios determined from the measurement of Strandberg {\it et al.}, in which no transition to the ground state from this resonance is observed \cite{PhysRevC.77.055801}. Given this discrepancy, it may be worth revisiting direct measurements of this resonance.

The resonance strength obtained in the experiment of Strandberg {\it et al.} \cite{PhysRevC.77.055801} is in good agreement with the result of Maas {\it et al.} \cite{MAAS1978213} for the same resonance. Therefore, despite the obvious inconsistencies in the nuclear data for this resonance, the resonance strength appears to be robust.

\subsection{The $E_x = 11.242$-MeV state}

An $E_x = 11.242$-MeV state which was assigned from a single $^{27}$Al($d,n$)$^{28}$Si measurement \cite{PhysRevC.32.394} is not observed in the $^{28}$Si($p,p^\prime$)$^{28}$Si measurement and we suggest that this state should be omitted from future compilations. There is no evidence from other $^{27}$Al($d.n$)$^{28}$Si experiments that this state exists \cite{PhysRevC.15.30}, nor from the $^{27}$Al($^3$He,$d$)$^{28}$Si reaction \cite{BOHNE1969273,NANN198261} which, as single-proton adding reactions, would be expected to populate this state.



\section{Evaluation of the $^{24}$Mg($\alpha,\gamma$)$^{28}$Si reaction rate}

In order to estimate the reaction rate with robust uncertainties, we used the publicly available Monte-Carlo code, {\tt RatesMC} \cite{0067-0049-207-1-18}. {\tt RatesMC} is available as part of the STARLIB project \cite{STARLIBWebsite} which calculates reaction rates using a Monte-Carlo technique. This approach allows for the evaluation of the reaction rate with meaningful uncertainties based on the uncertainties in the experimental data. For details as to how {\tt RatesMC} operates refer to Refs. \cite{Sallaska_2013,ILIADIS201031}. We briefly summarise here the assumptions made in preparing the inputs for {\tt RatesMC}, taking into account the remaining uncertainties in the nuclear data. 

For narrow resonances the reaction rate at a given temperature is determined from the resonance strengths and energies using \cite{iliadis2008nuclear}:

\begin{eqnarray}
    N_A \langle\sigma v\rangle = \frac{1.5399 \times 10^{11}}{(\mu T_9)^\frac{3}{2}} \sum_i (\omega\gamma)_i e^{-11.605 E_{r,i}/T_9},
\end{eqnarray}
given in cm$^3$ mol$^{-1}$ s$^{-1}$, where the sum is over the narrow resonances, $E_{r,i}$ is the resonance energy of the $i$th resonance, $(\omega\gamma)_i$ is the resonance strength of the $i$th resonance, $\mu$ is the reduced mass, and $T_9$ is the temperature in GK. For all of the resonances considered in this paper the $\alpha$-particle partial width is much smaller than the $\gamma$-ray partial width meaning that $\omega\gamma \approx (2J+1)\Gamma_\alpha$ where $J$ is the spin of the resonance and $\Gamma_\alpha$ is the $\alpha$-particle partial width. Therefore, the resonance strength is determined by the $\alpha$-particle partial width.

If a resonance strength was determined from thick-target measurements then the quantity $\omega\gamma$ is measured rather than $\omega$ and $\gamma$ separately as may happen in other cases. This product is insensitive to the spin and parity assignment of the underlying resonance; changes to spin-parity assignments will change both $\omega$ and $\gamma$ but will leave the total resonance strength and therefore the astrophysical reaction rate unchanged.

For cases where the spin and parity of a resonance are known but resonance strength or the partial widths are not, it is possible to estimate the possible contribution of the resonance to the reaction rate. The upper limit on the partial width may be calculated using the Wigner limit \cite{PhysRev.98.145} or some other single-particle width based on realistic wave-functions and these upper limits used to estimate the reaction rate. Various different approaches are available including calculating the rate assuming a mean value for the reduced width \cite{PhysRevC.77.055801}, calculating an upper limit based on upper limits for the reduced widths for cluster states \cite{lotay2019identification}, or, as in the present case, using a Monte-Carlo method where the reduced width is drawn from a probability distribution function \cite{0067-0049-207-1-18}.

In the present evaluation upper limits for resonance strengths have been calculated for all resonances below the lowest measured resonance ($E_r = 1010$ keV) in the experimental study of Strandberg {\it et al.} \cite{PhysRevC.77.055801}. For calculation of the reaction rates the limit for the resonance strength is the lower of (1) the limit from the Strandberg data ($\sum \omega\gamma < 2\times10^{-6}$ eV) for all resonances at $E_r \leq 909$ keV) or (2) the resonance strength assuming the Wigner limit for the $\alpha$-particle partial width. For the estimation of the contribution from the upper limits of resonances the {\tt RatesMC} code generates $\Gamma_\alpha$ partial widths assuming that the reduced widths follow a Porter-Thomas distribution with $\langle \theta^2\rangle = 0.01$ following systematic trends \cite{PhysRevC.85.065809}. We assume the same distribution for levels which have experimental upper limits, but use the experimental value as the upper limit instead of the Wigner limit.

The STARLIB evaluation of the $^{24}$Mg($\alpha,\gamma$)$^{28}$Si reaction rate did not include some resonances with unknown spin and parity (those at $E_x = 10.668$, $10.669$, $10.778$, $10.916$, $10.953$ and $11.101$ MeV). These represent one possible source of an increased reaction rate. In order to quantify the potential increase in the reaction rate from these resonances, we have evaluated the reaction rate including only those resonances in the STARLIB evaluation (denoted `Subset' in Fig. \ref{fig:RatioReactionRates}) and, additionally, including all isoscalar natural-parity resonances but omitting the $J^\pi = 3^-$ state at $E_x = 11.266$ MeV and any states known to be isovector in nature. This second reaction rate is denoted `All' in Fig. \ref{fig:RatioReactionRates}. The reaction rate evaluated with an increased set of resonances quantifies the potential increase in the reaction rate due to low-energy resonances. The state at $E_x = 11.266$ MeV is omitted as that resonance energy has been scanned in direct measurements in multiple experiments and no resonance has been observed. An experimental upper limit for the resonance strength based on the results of the direct $^{24}$Mg($\alpha,\gamma$)$^{28}$Si measurements would be preferable but is not possible. Table \ref{tab:Si28_nuclear_data} gives the nuclear data inputs for the resonances used for the calculation of the reaction rates.

We do not inflate the uncertainties in the resonance strengths to account for unknown systematic effects (as performed in Ref. \cite{PhysRevC.85.065809}) as, based on the analysis in Section \ref{sec:ConsistencyOfResonanceStrengths}, we find that the resonance strengths are in good agreement with each other.

Another potential source of systematic error is the use of atomic and nuclear masses in the computation of the resonance strengths \cite{PhysRevC.99.065809}. This leads to variations at 0.5 GK of around 5\%, smaller than the uncertainty in the reaction rate due to the uncertainties in the measurement of the resonance strengths. This effect is therefore within the uncertainty of our current evaluation.

\section{Calculated Reaction Rate}

\begin{figure}
\centering
\includegraphics[width=\columnwidth]{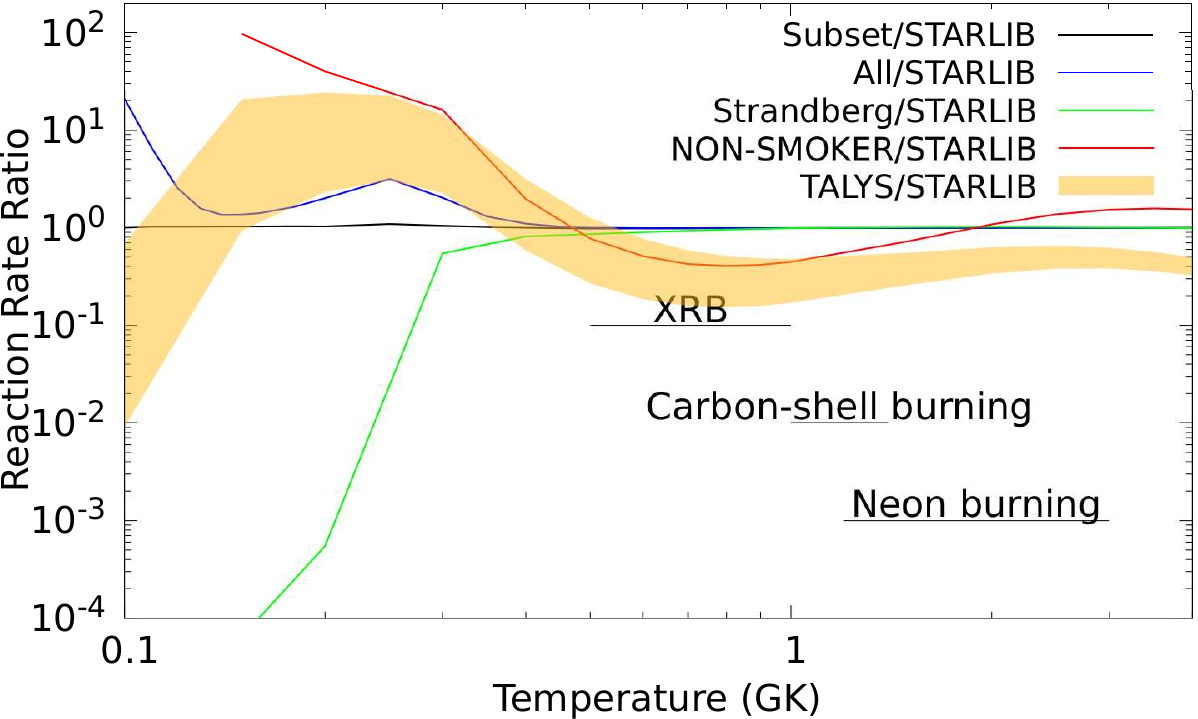}
\caption{Reaction-rate ratios relative to the STARLIB reaction rate for the reaction rates calculated in the present paper, the rate of Strandberg {\it et al.} \cite{PhysRevC.77.055801}, and reaction rates taken from the statistical-model codes TALYS \cite{TALYS} and NON-SMOKER \cite{RAUSCHER200147}. Reaction rates with TALYS were computed with the full range of $\alpha$-particle optical-model potentials with the band representing the most extreme values corresponding to the potentials of Avrigeanu {\it et al.} \cite{PhysRevC.90.044612} and the dispersive model of Demetriou, Grama and Goriely \cite{DEMETRIOU2002253}.}
 \label{fig:RatioReactionRates}
\end{figure}

The ratio of various reaction rates to the STARLIB reference rate are shown in Fig. \ref{fig:RatioReactionRates}. The reaction rates displayed include those calculated in the present study using both the STARLIB subset of levels and the complete set of levels, and the reaction rate of Strandberg {\it et al.} \cite{PhysRevC.77.055801}. Figure \ref{fig:RatioReactionRates} also includes a reaction rate calculated using TALYS (version 1.8) \cite{TALYS} with various $\alpha$ particle-nucleus optical-model potentials, as well as the reaction rate from the NON-SMOKER compilation \cite{RAUSCHER200147}.

The reaction rate calculated in the present paper using the STARLIB subset of states is consistent with the STARLIB evaluation. Above 0.4 GK, no significant variations in the reaction rate are observed depending on the choice of the nuclear data. Changes to the assumptions of the observed strength between $E_x =  11.14$ and $11.15$ MeV, or the spin-parity assignment of the level at $E_x = 10.806$ MeV have little effect on the final reaction rate. The addition of the low-lying resonances omitted in the STARLIB evaluation causes a modest increase in the reaction rate below 0.4 GK. The increase is comfortably below the temperatures of the astrophysical scenarios in which the $^{24}$Mg($\alpha,\gamma$)$^{28}$Si reaction plays a role.

The rates calculated in the present work and in Ref. \cite{0067-0049-207-1-18} are significantly higher at low temperatures than the Strandberg {\it et al.} \cite{PhysRevC.77.055801} rate. This is likely due to the different treatments of the $\alpha$-particle partial widths in the current Monte-Carlo calculations. However such low temperatures are not relevant to the astrophysical sites considered here. Meanwhile, the reaction rates from TALYS and NON-SMOKER over-predict the rate until around 0.4 GK. at which point both begin to under-predict the reaction rate. This is potentially due to the statistical models over-predicting other reaction channels such as $^{24}$Mg($\alpha,p$)$^{27}$Al or under-predicting the strength of decay $\gamma$-ray transitions.

\begin{figure}
 \includegraphics[width=\columnwidth]{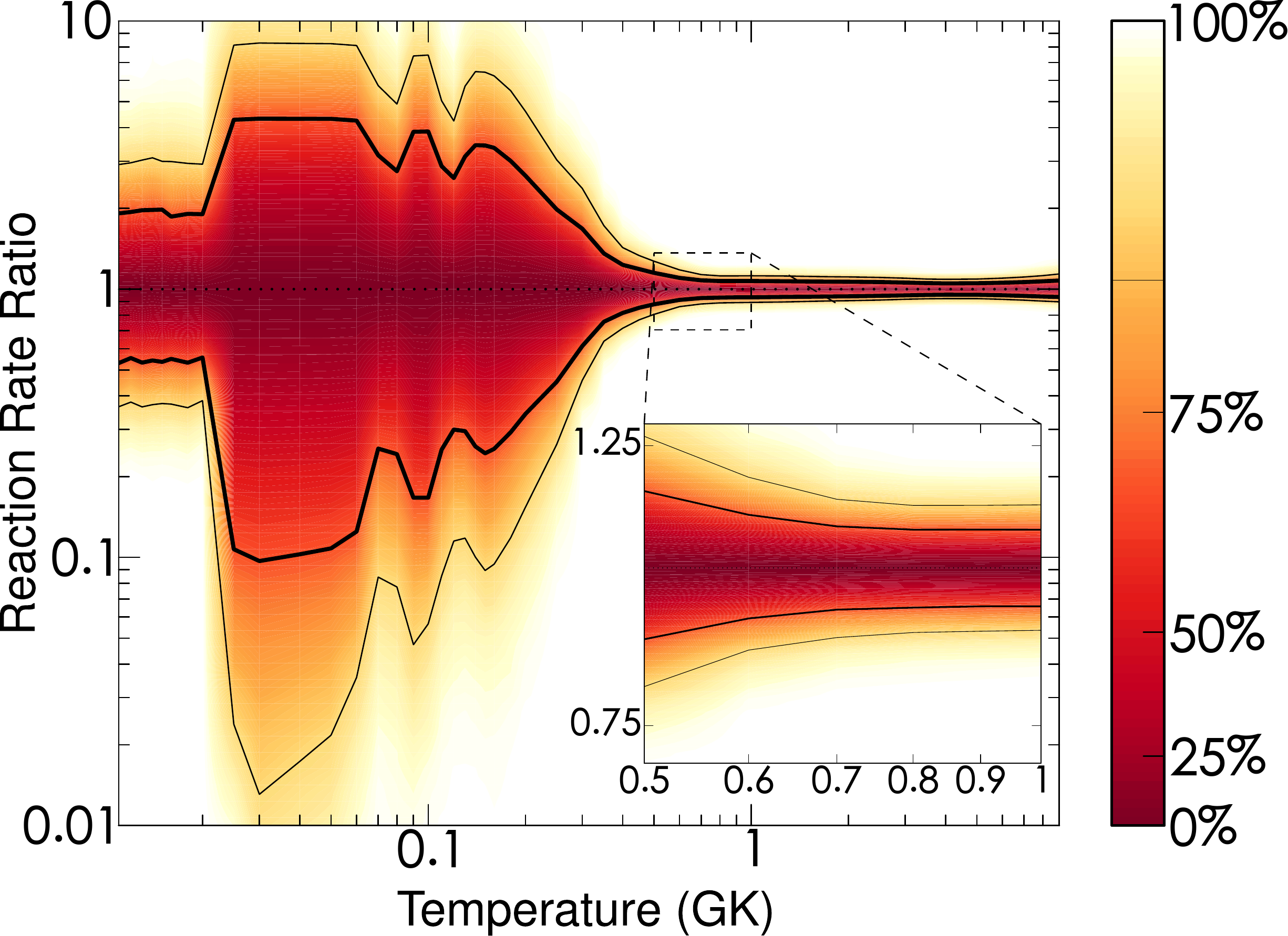}
 \caption{Reaction-rate uncertainty bands for scenario IV. The colour represents the probability distribution function while the 68\% and 95\% confidence limits are denoted by thick and thin black lines respectively.}
 \label{fig:ReactionRateUncertainty}
\end{figure}

\begin{figure}
    \centering
   \includegraphics[width=\columnwidth]{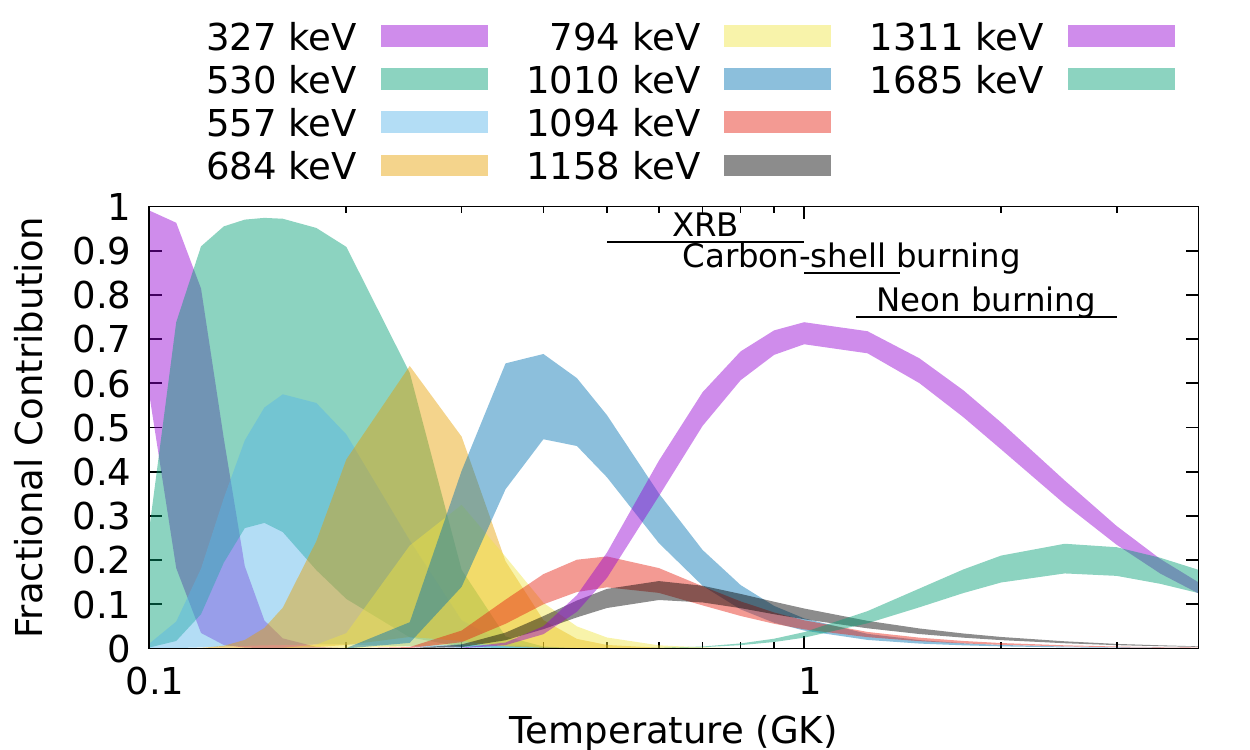}
    \caption{Fractional contribution plot of individual resonances to the $^{24}$Mg($\alpha,\gamma$)$^{28}$Si reaction rate, calculated using the complete set of possible additional resonances. Only resonances which contribute significantly to the reaction rate are plotted; the sum of the plotted contributions is below 100\% at some temperatures. The temperature regions corresponding to X-ray burst, carbon-shell burning and neon burning are also shown. The impact of potential new resonances ($E_r \leq 794$ keV) on the reaction rate is limited to temperatures below the astrophysically important temperature ranges.}
    \label{fig:FractionalContributions}
\end{figure}

The relative uncertainty in the $^{24}$Mg($\alpha,\gamma$)$^{28}$Si relative to the median reaction rate is shown in Fig. \ref{fig:ReactionRateUncertainty}. Above 0.4 GK the reaction rate is well constrained on the basis of the measured resonance strengths. Below 0.4 GK the reaction rate is dominated by unmeasured resonances which do not have known or estimated resonance strengths, in particular the $E_r = 197$-, $327$-, $530$-, $557$- and $684$-keV resonances (see Fig. \ref{fig:FractionalContributions}). The $\alpha$-particle partial widths are the dominant sources of uncertainty for the reaction rate below 0.4 GK.

The fractional contributions of individual resonances to the total reaction rate are shown in Fig. \ref{fig:FractionalContributions}. From Fig. \ref{fig:FractionalContributions}, additional resonances would have to lie above $E_r = 1010$ keV ($E_x = 10.994$ MeV) to cause increases to the reaction rate in the astrophysically important temperature region. Strong resonances above $E_r = 1010$ keV would have been observed in the numerous $^{24}$Mg($\alpha,\gamma$)$^{28}$Si direct measurements \cite{PhysRevC.77.055801,LYONS196925,MAAS1978213,SMULDERS19621093}. Large increases in the $^{24}$Mg($\alpha,\gamma$)$^{28}$Si reaction rate at astrophysically important temperatures due to unobserved resonances above $E_r = 1010$ keV can therefore be ruled out.

Tables \ref{tab:ScenarioIVRate} and \ref{tab:ScenarioVRate} provide the $^{24}$Mg($\alpha,\gamma$)$^{28}$Si median and 68\% confidence limit upper and lower limits reaction rates calculated using the STARLIB subset and the complete set of possible resonances, along with REACLIB parameterisations (Tables \ref{tab:REACLIBParametersScenarioIV} and \ref{tab:REACLIBParametersScenarioV}) of the reaction rate between 0.1 and 2.5 GK according to the prescription in Ref. \cite{Cyburt_2010}.

\begin{table*}
    \centering
    \caption{Reaction rate for the $^{24}$Mg($\alpha,\gamma$)$^{28}$Si reaction including only the subset of states included in the STARLIB compilation with updated nuclear data. The lower and upper limit reaction rates are the 68\% confidence limits.}
    \begin{ruledtabular}
    \begin{tabular}{c c c c}
    Temperature [GK] & Low Rate & Median Rate & High Rate \\ 
    \hline \\
    0.010 & 7.75$\times$10$^{-94}$ & 1.46$\times$10$^{-93}$ &
      2.84$\times$10$^{-93}$ \\ 
0.011 & 1.68$\times$10$^{-90}$ & 3.15$\times$10$^{-90}$ &
      6.01$\times$10$^{-90}$  \\ 
0.012 & 1.49$\times$10$^{-87}$ & 2.82$\times$10$^{-87}$ &
      5.35$\times$10$^{-87}$  \\ 
0.013 & 6.58$\times$10$^{-85}$ & 1.23$\times$10$^{-84}$ &
      2.36$\times$10$^{-84}$  \\ 
0.014 & 1.60$\times$10$^{-82}$ & 2.95$\times$10$^{-82}$ &
      5.68$\times$10$^{-82}$  \\ 
0.015 & 2.31$\times$10$^{-80}$ & 4.29$\times$10$^{-80}$ &
      8.35$\times$10$^{-80}$  \\ 
0.016 & 2.20$\times$10$^{-78}$ & 4.06$\times$10$^{-78}$ &
      7.70$\times$10$^{-78}$  \\ 
0.018 & 7.02$\times$10$^{-75}$ & 1.31$\times$10$^{-74}$ &
      2.52$\times$10$^{-74}$  \\ 
0.020 & 7.69$\times$10$^{-72}$ & 1.38$\times$10$^{-71}$ &
      2.68$\times$10$^{-71}$  \\ 
0.025 & 1.52$\times$10$^{-64}$ & 1.44$\times$10$^{-63}$ &
      5.99$\times$10$^{-63}$  \\ 
0.030 & 4.38$\times$10$^{-58}$ & 4.69$\times$10$^{-57}$ &
      1.96$\times$10$^{-56}$  \\ 
0.040 & 5.57$\times$10$^{-50}$ & 5.97$\times$10$^{-49}$ &
      2.50$\times$10$^{-48}$  \\ 
0.050 & 3.77$\times$10$^{-45}$ & 4.03$\times$10$^{-44}$ &
      1.69$\times$10$^{-43}$  \\ 
0.060 & 5.95$\times$10$^{-42}$ & 6.39$\times$10$^{-41}$ &
      2.68$\times$10$^{-40}$  \\ 
0.070 & 1.11$\times$10$^{-39}$ & 1.19$\times$10$^{-38}$ &
      4.98$\times$10$^{-38}$  \\ 
0.080 & 5.48$\times$10$^{-38}$ & 5.82$\times$10$^{-37}$ &
      2.43$\times$10$^{-36}$  \\ 
0.090 & 2.50$\times$10$^{-36}$ & 1.34$\times$10$^{-35}$ &
      5.05$\times$10$^{-35}$  \\ 
0.100 & 2.09$\times$10$^{-34}$ & 8.32$\times$10$^{-34}$ &
      2.54$\times$10$^{-33}$  \\ 
0.110 & 1.28$\times$10$^{-32}$ & 1.20$\times$10$^{-31}$ &
      5.09$\times$10$^{-31}$  \\ 
0.120 & 9.90$\times$10$^{-31}$ & 1.10$\times$10$^{-29}$ &
      4.70$\times$10$^{-29}$  \\ 
0.130 & 4.47$\times$10$^{-29}$ & 5.02$\times$10$^{-28}$ &
      2.15$\times$10$^{-27}$  \\ 
0.140 & 1.18$\times$10$^{-27}$ & 1.32$\times$10$^{-26}$ &
      5.65$\times$10$^{-26}$ \\ 
0.150 & 1.98$\times$10$^{-26}$ & 2.23$\times$10$^{-25}$ &
      9.51$\times$10$^{-25}$  \\ 
0.160 & 2.34$\times$10$^{-25}$ & 2.62$\times$10$^{-24}$ &
      1.12$\times$10$^{-23}$  \\ 
0.180 & 1.44$\times$10$^{-23}$ & 1.57$\times$10$^{-22}$ &
      6.73$\times$10$^{-22}$  \\ 
0.200 & 4.41$\times$10$^{-22}$ & 4.14$\times$10$^{-21}$ &
      1.75$\times$10$^{-20}$  \\ 
0.250 & 7.49$\times$10$^{-19}$ & 2.38$\times$10$^{-18}$ &
      6.89$\times$10$^{-18}$ \\ 
0.300 & 4.96$\times$10$^{-16}$ & 7.34$\times$10$^{-16}$ &
      1.22$\times$10$^{-15}$  \\ 
0.350 & 8.93$\times$10$^{-14}$ & 1.14$\times$10$^{-13}$ &
      1.52$\times$10$^{-13}$  \\ 
0.400 & 5.12$\times$10$^{-12}$ & 6.22$\times$10$^{-12}$ &
      7.65$\times$10$^{-12}$  \\ 
0.450 & 1.29$\times$10$^{-10}$ & 1.51$\times$10$^{-10}$ &
      1.79$\times$10$^{-10}$  \\ 
0.500 & 1.83$\times$10$^{-09}$ & 2.09$\times$10$^{-09}$ &
      2.40$\times$10$^{-09}$  \\ 
0.600 & 1.14$\times$10$^{-07}$ & 1.25$\times$10$^{-07}$ &
      1.38$\times$10$^{-07}$  \\ 
0.700 & 2.45$\times$10$^{-06}$ & 2.65$\times$10$^{-06}$ &
      2.86$\times$10$^{-06}$  \\ 
0.800 & 2.59$\times$10$^{-05}$ & 2.78$\times$10$^{-05}$ &
      2.98$\times$10$^{-05}$  \\ 
0.900 & 1.66$\times$10$^{-04}$ & 1.78$\times$10$^{-04}$ &
      1.91$\times$10$^{-04}$  \\ 
1.000 & 7.47$\times$10$^{-04}$ & 8.00$\times$10$^{-04}$ &
      8.58$\times$10$^{-04}$  \\ 
1.250 & 1.16$\times$10$^{-02}$ & 1.24$\times$10$^{-02}$ &
      1.32$\times$10$^{-02}$  \\ 

    \end{tabular}
    \end{ruledtabular}
    \label{tab:ScenarioIVRate}
\end{table*}

\begin{table}[th]
    \centering
    \caption{REACLIB parameters for the reaction rate given in Table \ref{tab:ScenarioIVRate} corresponding to the STARLIB subset of states.}
    \begin{ruledtabular}
    \begin{tabular}{c c c c}
       Parameter & Low & Median & High \\
       \hline \\
       $a_0$ & $ 3558 $ & $ 3923 $ & $ 3480 $ \\
$a_1$ & $ -50.52 $ & $ -67.83 $ & $ -67.2 $ \\
$a_2$ & $ 3099 $ & $ 3880 $ & $ 3696 $ \\
$a_3$ & $ -7245 $ & $ -8446 $ & $ -7738 $ \\
$a_4$ & $ 706.3 $ & $ 782.8 $ & $ 691 $ \\
$a_5$ & $ -74.76 $ & $ -79.65 $ & $ -67.97 $ \\
$a_6$ & $ 2829 $ & $ 3404 $ & $ 3180 $
    \end{tabular}
    \end{ruledtabular}
    \label{tab:REACLIBParametersScenarioIV}
\end{table}

\begin{table*}
    \centering
    \caption{Reaction rate for the $^{24}$Mg($\alpha,\gamma$)$^{28}$Si reaction including all possible contributing states. The lower and upper limit reaction rates are the 68\% confidence limits.}
    \begin{ruledtabular}
    \begin{tabular}{c c c c}
        Temperature [GK] & Low Rate & Median Rate & High Rate \\
        \hline \\
        
    0.010 & 7.91$\times$10$^{-94}$ & 1.46$\times$10$^{-93}$ &
      2.82$\times$10$^{-93}$ \\ 
0.011 & 1.69$\times$10$^{-90}$ & 3.20$\times$10$^{-90}$ &
      6.13$\times$10$^{-90}$ \\ 
0.012 & 1.53$\times$10$^{-87}$ & 2.76$\times$10$^{-87}$ &
      5.35$\times$10$^{-87}$\\ 
0.013 & 6.51$\times$10$^{-85}$ & 1.22$\times$10$^{-84}$ &
      2.41$\times$10$^{-84}$  \\ 
0.014 & 1.59$\times$10$^{-82}$ & 2.93$\times$10$^{-82}$ &
      5.79$\times$10$^{-82}$  \\ 
0.015 & 2.25$\times$10$^{-80}$ & 4.21$\times$10$^{-80}$ &
      8.33$\times$10$^{-80}$ \\ 
0.016 & 2.24$\times$10$^{-78}$ & 4.07$\times$10$^{-78}$ &
      7.59$\times$10$^{-78}$  \\ 
0.018 & 6.87$\times$10$^{-75}$ & 1.29$\times$10$^{-74}$ &
      2.46$\times$10$^{-74}$ \\ 
0.020 & 7.83$\times$10$^{-72}$ & 1.41$\times$10$^{-71}$ &
      2.68$\times$10$^{-71}$  \\ 
0.025 & 1.52$\times$10$^{-64}$ & 1.42$\times$10$^{-63}$ &
      6.08$\times$10$^{-63}$  \\ 
0.030 & 4.43$\times$10$^{-58}$ & 4.58$\times$10$^{-57}$ &
      1.98$\times$10$^{-56}$  \\ 
0.040 & 6.01$\times$10$^{-50}$ & 5.85$\times$10$^{-49}$ &
      2.53$\times$10$^{-48}$  \\ 
0.050 & 4.30$\times$10$^{-45}$ & 3.98$\times$10$^{-44}$ &
      1.72$\times$10$^{-43}$  \\ 
0.060 & 7.97$\times$10$^{-42}$ & 6.39$\times$10$^{-41}$ &
      2.71$\times$10$^{-40}$ \\ 
0.070 & 4.54$\times$10$^{-39}$ & 1.78$\times$10$^{-38}$ &
      5.63$\times$10$^{-38}$  \\ 
0.080 & 7.67$\times$10$^{-37}$ & 3.16$\times$10$^{-36}$ &
      8.71$\times$10$^{-36}$  \\ 
0.090 & 5.06$\times$10$^{-35}$ & 3.04$\times$10$^{-34}$ &
      1.17$\times$10$^{-33}$  \\ 
0.100 & 2.91$\times$10$^{-33}$ & 1.74$\times$10$^{-32}$ &
      6.75$\times$10$^{-32}$  \\ 
0.110 & 1.91$\times$10$^{-31}$ & 7.59$\times$10$^{-31}$ &
      2.18$\times$10$^{-30}$  \\ 
0.120 & 8.20$\times$10$^{-30}$ & 2.74$\times$10$^{-29}$ &
      7.11$\times$10$^{-29}$  \\ 
0.130 & 2.27$\times$10$^{-28}$ & 7.68$\times$10$^{-28}$ &
      2.40$\times$10$^{-27}$  \\ 
0.140 & 4.51$\times$10$^{-27}$ & 1.74$\times$10$^{-26}$ &
      5.99$\times$10$^{-26}$  \\ 
0.150 & 7.20$\times$10$^{-26}$ & 2.94$\times$10$^{-25}$ &
      1.01$\times$10$^{-24}$  \\ 
0.160 & 9.11$\times$10$^{-25}$ & 3.59$\times$10$^{-24}$ &
      1.21$\times$10$^{-23}$  \\ 
0.180 & 7.31$\times$10$^{-23}$ & 2.50$\times$10$^{-22}$ &
      7.51$\times$10$^{-22}$  \\ 
0.200 & 2.76$\times$10$^{-21}$ & 8.06$\times$10$^{-21}$ &
      2.13$\times$10$^{-20}$  \\ 
0.250 & 3.12$\times$10$^{-18}$ & 6.90$\times$10$^{-18}$ &
      1.37$\times$10$^{-17}$ \\ 
0.300 & 8.75$\times$10$^{-16}$ & 1.43$\times$10$^{-15}$ &
      2.39$\times$10$^{-15}$  \\ 
0.350 & 1.10$\times$10$^{-13}$ & 1.46$\times$10$^{-13}$ &
      1.98$\times$10$^{-13}$ \\ 
0.400 & 5.57$\times$10$^{-12}$ & 6.83$\times$10$^{-12}$ &
      8.42$\times$10$^{-12}$  \\ 
0.450 & 1.34$\times$10$^{-10}$ & 1.57$\times$10$^{-10}$ &
      1.87$\times$10$^{-10}$ \\ 
0.500 & 1.87$\times$10$^{-09}$ & 2.13$\times$10$^{-09}$ &
      2.45$\times$10$^{-09}$  \\ 
0.600 & 1.15$\times$10$^{-07}$ & 1.26$\times$10$^{-07}$ &
      1.39$\times$10$^{-07}$  \\ 
0.700 & 2.46$\times$10$^{-06}$ & 2.66$\times$10$^{-06}$ &
      2.87$\times$10$^{-06}$  \\ 
0.800 & 2.59$\times$10$^{-05}$ & 2.79$\times$10$^{-05}$ &
      2.99$\times$10$^{-05}$  \\ 
0.900 & 1.66$\times$10$^{-04}$ & 1.79$\times$10$^{-04}$ &
      1.91$\times$10$^{-04}$  \\ 
1.000 & 7.47$\times$10$^{-04}$ & 8.01$\times$10$^{-04}$ &
      8.59$\times$10$^{-04}$  \\ 
1.250 & 1.16$\times$10$^{-02}$ & 1.24$\times$10$^{-02}$ &
      1.32$\times$10$^{-02}$  \\ 
    \end{tabular}
    \end{ruledtabular}
    \label{tab:ScenarioVRate}
\end{table*}

\begin{table}[th]
    \centering
    \caption{REACLIB parameters for the reaction rate given in Table \ref{tab:ScenarioVRate} corresponding to the complete set of states.}
    \begin{ruledtabular}
    \begin{tabular}{c c c c}
       Parameter & Low & Median & High \\
       \hline \\
       $a_0$ & $ 1495 $ & $ 648.1 $ & $ 52.49 $ \\
$a_1$ & $ -22.61 $ & $ -13.38 $ & $ -7.904 $ \\
$a_2$ & $ 1326 $ & $ 682.7 $ & $ 265 $ \\
$a_3$ & $ -3053 $ & $ -1415 $ & $ -297.5 $ \\
$a_4$ & $ 273.6 $ & $ 96.19 $ & $ -27.91 $ \\
$a_5$ & $ -26.21 $ & $ -5.648 $ & $ 8.922 $ \\
$a_6$ & $ 1221 $ & $ 612.9 $ & $ 206.7 $
    \end{tabular}
    \end{ruledtabular}
    \label{tab:REACLIBParametersScenarioV}
\end{table}

\begin{figure}
 \includegraphics[width=\columnwidth]{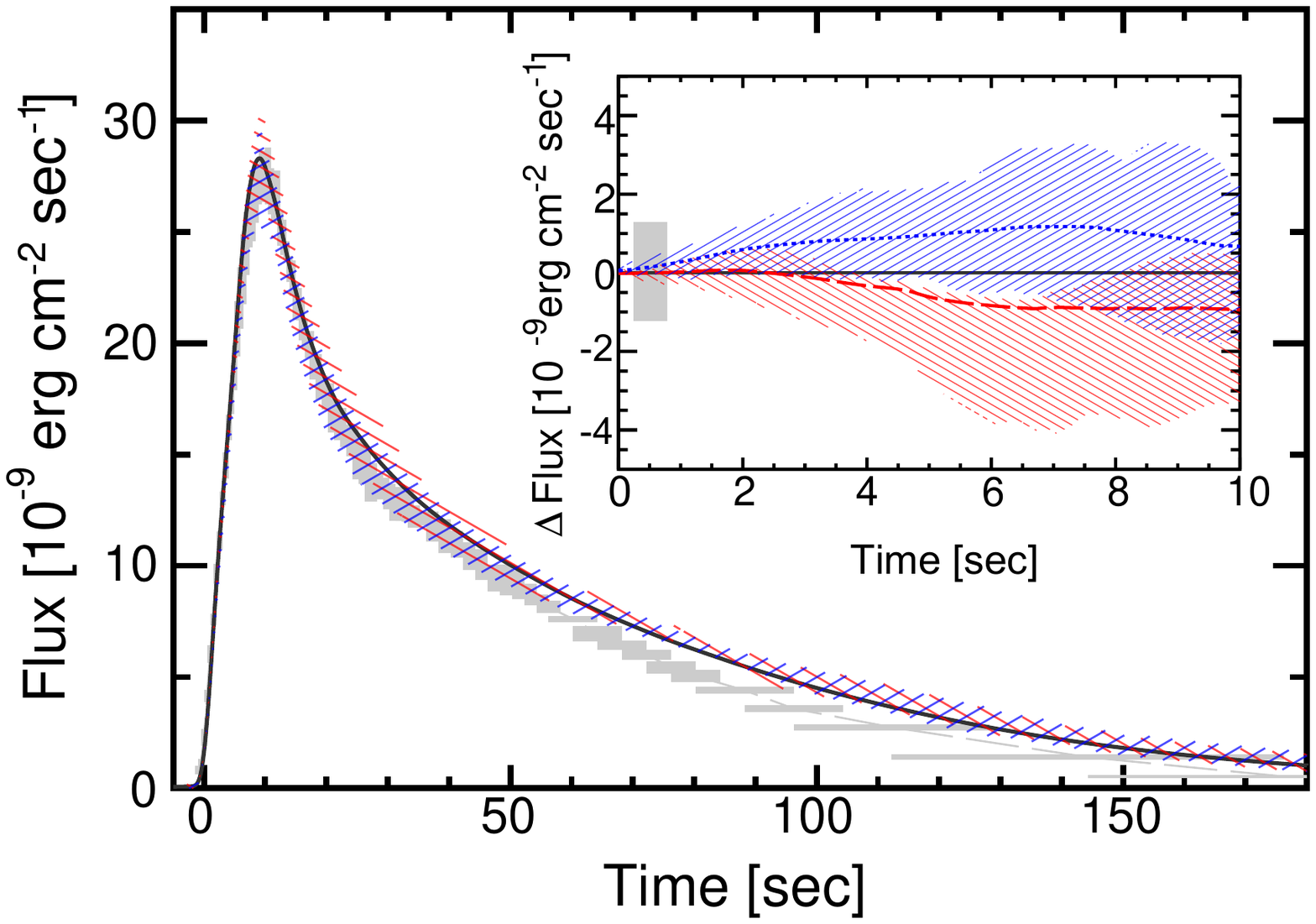}
 \caption{X-ray burst light curve 68\% confidence interval bands calculated with {\tt MESA} using the median (black line), upper 95\% (red-hash area), and lower 95\% (blue-hash area) STARLIB reaction rates for $^{24}{\rm Mg}(\alpha,\gamma)^{28}{\rm Si}$, compared to observations of the year 2007 bursting epoch of GS 1826-24 (gray boxes) for context. The inset shows the residual over the light curve rise to {\tt MESA} results obtained with the median STARLIB rate, where the gray box indicates the average observational uncertainty in that time frame. Note that the {\tt MESA} light curve bands are asymmetric uncertainties, where the dashed or dotted line indicates the average, with the uncertainty for a single band owing to burst-to-burst variability.}
 \label{fig:XRBlc}
\end{figure}

\section{Impact of reaction-rate uncertainty on the X-ray burst lightcurve}

The re-analysed $^{24}$Mg($\alpha,\gamma$)$^{28}$Si reaction rate was used in the model of Refs.~\cite{Meisel_2018,Meisel_2019} to determine if the remaining uncertainty in the reaction rate causes any discernible variation in the behaviour of the X-ray burst lightcurve. One-dimensional model calculations were performed with the code {\tt MESA}~\cite{Paxton2011,Paxton2013,Paxton2015} version 9793, following the thermodynamic and nucleosynthetic evolution of an $\sim0.01$~km thick envelope of material discretized into $\sim1000$~zones with an inner boundary of an 11.2~km 1.4~$M_{\odot}$ neutron star. Notable microphysics included time-dependent mixing-length theory for convection~\cite{Henyey1969}, a post-Newtonian correction to local gravity for general relativistic effects~\cite{Paxton2011}, and the 304~isotope network of Ref.~\cite{Fisker2008} employing the REACLIBv2.2 nuclear reaction rate library~\cite{Cyburt_2010}. The accretion conditions used were those found by Ref.~\cite{Meisel_2018} to best reproduce the observed features of the year 2007 bursting epoch of the source GS 1826-24~\cite{Galloway2017}.

Average light curves were calculated from a sequence of 14 X-ray bursts, employing either the median, upper 95\% confidence limit, or lower 95\% confidence limit for the $^{24}{\rm Mg}(\alpha,\gamma)^{28}{\rm Si}$ reaction rate calculated using {\tt RatesMC}. Results are shown in Fig.~\ref{fig:XRBlc} alongside observational data~\cite{Galloway2017} for GS 1826-24 for context. The figure inset shows the residual between calculation results using the median rate and upper and lower 95\% confidence intervals for the rising portion of the X-ray burst light curve. Whereas the previously assumed factor of 10 uncertainty resulted in an appreciably different convexity of the light curve rise~\cite{Meisel_2019}, the present rate uncertainty causes light curve variations on the order of observational uncertainties and the intrinsic burst-to-burst variability of model calculations. Therefore, we find the $^{24}{\rm Mg}(\alpha,\gamma)^{28}{\rm Si}$ reaction rate uncertainty no longer appreciably contributes to the overall uncertainty in the calculated X-ray burst light curve.

\section{Conclusions}

In summary, the available nuclear data on $^{28}$Si relevant to the $^{24}$Mg($\alpha,\gamma$)$^{28}$Si reaction have been reviewed and the reaction rate recalculated on the basis of new level assignments. At astrophysically important temperatures the reaction rate is dominated by the $E_r = 1010$, $1094$, $1164$, and $1311$ keV resonances, with the last of these dominating the rate in the astrophysically relevant temperature range. Direct measurements of $^{24}$Mg($\alpha,\gamma$)$^{28}$Si resonance strengths are consistent and the level of uncertainty in the calculated reaction rate is small. The direct measurement data of Refs. \cite{PhysRevC.77.055801,MAAS1978213,LYONS196925,SMULDERS19621093} include excitation functions over a range of incident energies, and therefore any unobserved resonances cannot be sufficiently strong to significantly increase the $^{24}$Mg($\alpha,\gamma$)$^{28}$Si reaction rate. Additional contributions from new states have been ruled out using $^{28}$Si($p,p^\prime$)$^{28}$Si data. This unselective reaction revealed no new states between the $\alpha$-particle threshold and the lowest directly measured resonance.

The remaining uncertainty in the $^{24}{\rm Mg}(\alpha,\gamma)^{28}{\rm Si}$ reaction rate does not significantly alter the calculated X-ray burst light curve. Therefore, this reaction rate, one of eight identified as important for X-ray burst model-observation comparisons~\cite{Meisel_2019}, is adequately constrained for the purposes of determining properties of accreting neutron star systems from such comparisons.

\section*{Acknowledgements}

This paper is based upon work from the \textquoteleft ChETEC\textquoteright\ COST Action (CA16117), supported by COST (European Cooperation in Science and Technology). PA acknowledges and thanks the Claude Leon Foundation for a postdoctoral research fellowship and the NRF/iThemba LABS for support, and those who assisted in the collection of the $^{28}$Si($\alpha,\alpha^\prime$)$^{28}$Si data using the K600 magnetic spectrometer at iThemba LABS and $^{28}$Si($p,p^\prime$)$^{28}$Si data using the Q3D magnetic spectrograph at the Maier-Leibnitz Laboratorium in Munich. AML acknowledges the support of the U.K. Science and Technology Facilities Council (STFC Consolidated Grant ST/P003885/1).
ZM acknowledges support from the U.S. Department of Energy Office of Science through Grants No. DE-FG02-88ER40387 and DESC0019042 and the U.S. National Science Foundation through Grant PHY-1430152 (Joint Institute for Nuclear Astrophysics -- Center for the Evolution of the Elements). The authors thank Richard Longland for access to the {\tt RatesMC} code and associated R scripts. The authors thank Joachim G\"{o}rres for useful information about the $^{24}$Mg($\alpha,\gamma$)$^{28}$Si experiment of Strandberg {\it et al.} \cite{PhysRevC.77.055801} and noting the strength of the transition from the ground state in the $^{28}$Si($e,e^\prime$)$^{28}$Si data, and Peter von Neumann-Cosel for useful comments about distinguishing $E0$ and $E2$ transitions with electron scattering.

\bibliography{Mg24AlphaGammaReanalysisPaper}

\end{document}